\documentclass[preprintnumbers,amsmath,amssymb,superscriptaddress]{revtex4}
\usepackage{epsfig,bm}
\usepackage{ulem}
\usepackage{amsmath,color}
\usepackage{graphics} 
\usepackage{amssymb}

\begin{document}

\title{Adiabatic hyperspherical approach to large-scale nuclear dynamics}

\author{Y. Suzuki}
\affiliation{Department of Physics, Niigata University, Niigata 950-2181, Japan}
\affiliation{RIKEN Nishina Center, Wako 351-0198, Japan}

\begin{abstract}%
We formulate a fully microscopic approach to large-scale nuclear dynamics using a hyperradius 
as a collective coordinate. An adiabatic potential is defined by taking account of all possible configurations at a fixed hyperradius, and its hyperradius 
dependence plays a key role in governing 
the global nuclear motion. In order to go to larger systems beyond few-body systems, we suggest basis functions of a microscopic multicluster model, propose a method for calculating matrix elements of an adiabatic Hamiltonian with use of Fourier transforms, and test its effectiveness. 
\end{abstract}

\maketitle

\section{Introduction}
\label{intro}

Atomic nuclei present a unique example of self-bound, finite quantum many-body systems. 
They not only exhibit a variety of excitation modes but also decay or fission into two or 
a few fragments. Exploring the excitation mechanism based on 
single-particle, collective and clustering degrees of freedom is an interesting   
subject. Intrinsically different shapes such as prolate-oblate may coexist or mix at close energies, leading 
to the so-called large-amplitude collective motion~\cite{matsuyanagi10}. Spontaneous fission and sub-barrier 
fusion are also typical examples of the collective motion that 
involves a large-scale change of the nuclear size~\cite{krappe12,hagino12}. A fully microscopic description of  their dynamics is still a long-standing challenging problem.

All of the above phenomena should in principle be described 
starting from a Hamiltonian of the system. 
What is often performed is, however, to solve an equation of motion with some 
constraints~\cite{ring80} or to calculate 
energy surfaces assuming different shapes in order to look for a path along which the collective 
motion proceeds. 
In the case of a deep sub-barrier fusion an initial fragment 
decomposition is maintained for the whole fusion process and the relevant fusion potential is 
calculated as a function of the relative distance of the fragments. It is hard for that approach to take into account 
couplings with configurations corresponding to a different mass distribution of the fragments.
Since the phenomena are very complicated, those approaches 
sound reasonable. However, neither the geometrical shape nor the relative distance between the fragments is a nuclear collective coordinate in a strict sense. 
A question thus arises of whether or not we can describe the large-scale dynamics by employing a true collective coordinate. 

The purpose of this paper is to make use of a hyperradius as a collective coordinate, 
and to step forward for a consistent formulation of the large-scale dynamics together  
with the underlying collective potential. Most of the needed ingredients are available in the  literature. 
The hyperradius is a global coordinate that measures matter size, and it is 
widely used in three-body problems~\cite{zhukov93,lin95,krivec98,nielsen01}.  
After pioneering work with the hyperspherical approach~\cite{macek68}, its 
extension to $N$-body systems has been proposed for solving various 
problems~\cite{barnea01,timofeyuk08,mehta09,rakshit12,daily14}. A common foundation of 
all the hyperspherical approaches is that a total wave function of the system 
is expanded in terms of 
a product of the hyperradial and hyperangular functions. There are two types of 
realization for describing the hyperangular functions. One is to use hyperspherical 
harmonics~\cite{barnea01,timofeyuk08}, and the other called an adiabatic hyperspherical 
approach is to employ channel wave functions 
that are defined by diagonalizing the hyperangular part of the Hamiltonian~\cite{rakshit12,daily14}. The former has the advantage that the hyperspherical 
harmonics are well-known eigenfunctions of the angular part of 
the multi-dimensional Laplacian, but its use in a real problem 
is fairly complicated and so far limited to few-particle systems. Moreover, 
a convergence with that expansion is rather slow. 
See the first paper in Ref.~\cite{timofeyuk08} for details on the development and difficulty. 
The latter is widely used in atomic and molecular physics. Since an adiabatic hyperspherical potential defined 
there reflects the large-scale change of the system, we adopt the adiabatic hyperspherical approach in what follows.  

The equation of motion in the hyperspherical method 
is the same independent of the number of particles in the system, which is 
an appealing feature of the hyperspherical approach. In spite of the various 
efforts, only small systems have so far been investigated mainly because 
calculating the matrix element  
of an adiabatic Hamiltonian is still not trivial and solving its eigenvalue problem is 
hard for general $N$-body systems. Correlated Gauss functions (CG) are employed for studying cold atom physics 
and electron-positron systems in Refs.~\cite{rakshit12,daily14}, but 
their application is limited to few-body systems with the total orbital angular momentum $L=0$ and 1. 
The use of harmonic-oscillator shell-model wave functions 
is discussed in Ref.~\cite{timofeyuk08} for calculating the matrix element needed in the hyperspherical approach. The  oscillator basis is convenient for representing such one-centered configurations that are not highly excited from the ground state, 
but it is not flexible enough to cope with a description of large-scale change such as,  
for example, the clustering and fragmentation. 
We instead need basis functions of cluster type to describe such configurations. We attempt 
here an 
extension of microscopic multicluster wave functions~\cite{varga94,arai01} used to describe the structure of light nuclei. 
In the multicluster model the intrinsic fragment wave functions are described with shell-model type configurations, while the relative motion among the fragments is described with the CG~\cite{varga95,book}. 
We apply a Fourier integral for evaluating the matrix element as it is applicable  
for any type of many-body basis function. 

The structure of the present paper is as follows. We define in Sect.~\ref{HH.coord} the hyperspherical coordinates and 
separate the kinetic energy of the system into hyperradial and hyperangular parts. 
In Sect.~\ref{eq.Motion} we define the eigenvalue problem of the adiabatic Hamiltonian 
and present the equation of motion for hyperradial functions. In Sect.~\ref{rho-dep.pot} we discuss qualitative features of the adiabatic potential together with a separation of active and inactive degrees of freedom. 
In Sect.~\ref{Matrix.element} we define basis functions of the multicluster model and give a method for 
calculating the matrix element integrated over the hyperangles together with 
examples for the overlap and kinetic energy. 
In Sect.~\ref{sec.shape} we show how to extract the evolution of intrinsic shapes 
of the system 
as a function of the hyperradius. In Sect.~\ref{sec.constraint} we touch on 
an eigenvalue problem of the full Hamiltonian 
with a constraint of the mean-square matter radius in comparison with the present approach. Conclusions are drawn in Sect.~\ref{summary}.

\section{Hyperspherical coordinates}
\label{HH.coord}
 
We start from defining the hyperradius for a general case consisting of $K$ particles. 
The mass of the $i$th particle is $A_i$ in units of a suitable  
mass $m$. By denoting its position coordinate by $\bm R_i$,   
we define a set of Jacobi coordinates by
\begin{align}
\bm X_i=\sqrt{\mu_i}
\Big(\bm R_{i+1}-\frac{1}{A_{12\ldots i}}\sum_{j=1}^i A_j{\bm R}_j\Big),
\label{XtoR}
\end{align}
where $A_{12 \ldots i}=\sum_{j=1}^i A_j$ and $\mu_i$ is the reduced mass factor, 
$\mu_i=A_{12\ldots i}A_{i+1}/A_{12\ldots {i+1}}$.
The square of the hyperradius $\rho$ is defined by 
\begin{align}
\rho^2= \sum_{i=1}^{K-1}\bm X_i^2,
\label{rhosq.N-body}
\end{align}
which is also rewritten in several ways as 
\begin{align}
\rho^2=\sum_{i=1}^K A_i(\bm R_i-\bm R_{\rm cm})^2=\sum_{i=1}^KA_i\bm R_i^2-A_{12\ldots K}\bm R_{\rm cm}^2=\frac{1}{A_{12\ldots K}}\sum_{j>i=1}^KA_iA_j(\bm R_i-\bm R_j)^2, 
\label{N-body}
\end{align}
where $\bm R_{\rm cm}$ is the center-of-mass (cm) coordinate of the system, $\bm R_{\rm cm}=\sum_{i=1}^K A_i\bm R_i/A_{12\ldots K}$. Note that $m\rho^2$ is equal to the trace of the moment of inertia tensor of the system. 

It is straightforward to extend the above definition to an $N$-nucleon system. 
Protons and neutrons are assumed to have an equal mass, the nucleon mass, which is taken 
as $m$. By denoting the nucleon's position coordinate by $\bm r_i$, we define 
Jacobi coordinates as   
\begin{align}
\bm x_i=\sqrt{\mu_i}\Big(\bm r_{i+1}-\frac{1}{i}\sum_{j=1}^i\bm r_j\Big) 
\label{r.to.x}
\end{align}
with $\mu_i=i/(i+1)$. Then $\rho^2$ reads  
\begin{align}
\rho^2=\sum_{i=1}^{N-1}\bm x_i^2
=\sum_{i=1}^N(\bm r_i-\bm R_{\rm cm})^2=
\sum_{i=1}^N\bm r_i^2-N\bm R_{\rm cm}^2=\frac{1}{N}\sum_{j>i=1}^N(\bm r_i-\bm r_j)^2. 
\label{rho2.xx}
\end{align}

We often use a matrix notation. For example, 
$\bm x=(\bm x_i)$ stands for an $N-1$-dimensional 
column vector or an $(N-1)\times 1$ matrix, and $\tilde{\bm x}$ stands for its row vector.  
The $\rho^2$ is simply written as a scalar product,  
$\rho^2=\tilde{\bm x}\bm x$. It is clear that $\rho^2$ is equally defined 
by any coordinates that are related to $\bm x$ by  
an orthogonal transformation. In fact $\rho^2$ is independent of any choice of such coordinates.

A measure of the nuclear size, $\rho^2/N$ is an operator for the mean-square matter radius. 
Symmetric with respect to the nucleons' coordinates, $\rho$ is a collective coordinate 
that has a unit of length. 
The other $3N-4$ coordinates are hyperangle coordinates 
denoted by $\Omega$ collectively. The volume element for integration reads 
\begin{align}
d\bm x=d\bm x_1d\bm x_2 \ldots d\bm x_{N-1}=\rho^{d-1}d\rho d\Omega, 
\label{vol.elem}
\end{align}
where $d$ is the dimension of the spatial coordinates excluding the cm coordinate
\begin{align}
d=3(N-1).
\label{dimension}
\end{align}
It is well known that the volume $V_d$ of a $d$-dimensional hypersphere with radius $\rho=\rho_0$ is given by 
$V_d\equiv \int_{\tilde{\bm x}\bm x \leq \rho_0^2} d\bm x=(\rho_0{\sqrt{\pi}})^d/\Gamma({d}/{2}+1)$ with the gamma 
function $\Gamma$. Since $V_d$ 
is equal to $\int_0^{\rho_0} \rho^{d-1}d\rho \int d\Omega$, the surface area of the hypersphere is  
\begin{align}
\int d\Omega=\frac{2\sqrt{\pi}^{d}}{\Gamma({d}/{2})}.
\label{surface.integral}
\end{align}
The volume element in the single-particle coordinates reads $d\bm r_1d\bm r_2 \ldots d\bm r_{N}=N^{3/2}d\bm x d\bm R_{\rm cm}$. 

Let us introduce dimensionless coordinates  
$\bm \xi_i$ by $\bm x_i=\rho\bm \xi_i$.  They are subject to the constraint  
$\sum_{i=1}^{N-1}\bm \xi_i^2=\tilde{\bm \xi}\bm \xi=1$.  An explicit form of $\Omega$ 
may be constructed from the $N-1$ 
vectors $\bm \xi_i$, but it is not needed in what follows. 
It should be noted, however, that a variety of 
configurations or shapes of the nucleus correspond to different functions of $\Omega$. 
The total kinetic energy $T$ of the $N$-nucleon system, with its cm kinetic energy 
$T_{\rm cm}$ being subtracted,  
is separated into hyperradial ($T_{\rho}$) and hyperangular ($T_{\Omega}$) parts:  
\begin{align}
T&=-\frac{\hbar^2}{2m}\sum_{i=1}^{N}\frac{\partial^2}{\partial \bm r_i^2}- T_{\rm cm} 
%\notag \\
%&=-\frac{\hbar^2}{2m}\sum_{i=1}^{N-1}\frac{\partial^2}{\partial \bm x_i^2}\notag \\
%&=T_{\rho}+T_{\Omega}
=-\frac{\hbar^2}{2m}\sum_{i=1}^{N-1}\frac{\partial^2}{\partial \bm x_i^2}
=T_{\rho}+T_{\Omega}
\label{kinetic.op}
\end{align}
with 
\begin{align}
T_{\rho}&=-\frac{\hbar^2}{2m} \left(\frac{\partial^2}{\partial \rho^2}+\frac{d-1}{\rho}\frac{\partial}{\partial \rho}\right)
%\notag \\
%&=-\frac{\hbar^2}{2m}\rho^{-(d-1)}\frac{\partial}{\partial \rho}\rho^{d-1}\frac{\partial}{\partial \rho}.
=-\frac{\hbar^2}{2m}\rho^{-(d-1)}\frac{\partial}{\partial \rho}\rho^{d-1}\frac{\partial}{\partial \rho}.
\end{align}
The hyperangular kinetic energy $T_{\Omega}$ may be expressed as  
\begin{align}
T_{\Omega}=\frac{\hbar^2 {\cal K}(\Omega)^2}{2m \rho^2},
\end{align}
where ${\cal K}(\Omega)^2$ is the square of the grand angular momentum. An explicit form of ${\cal K}(\Omega)^2$ is available in a recursive way together 
with the definition of $\Omega$~\cite{lin95,rakshit12}.

Suppose that the $N$-nucleon system develops into $K$ fragments or clusters each of which has $N_i$ nucleons ($\sum_{i=1}^K N_i=N$). It is convenient to divide $\rho^2$ of 
Eq.~(\ref{rho2.xx}) into two groups: 
\begin{align}
\rho^2=\rho^2_{\rm in}+\rho^2_{\rm rel}=\sum_{i=1}^{K}\rho_i^2+\sum_{i=1}^KN_i(\bm R_i-\bm R_{\rm cm})^2,
\label{rho.2frag}
\end{align}
where $\bm R_i$ is the cm coordinate of the $i$th fragment and $\rho_i^2$ is its squared 
hyperradius,
\begin{align}
\rho_i^2 =\sum_{j=1}^{N_i}(\bm r_{N_{12 \ldots i-1}+j}-\bm R_i)^2,\ \ \ \ \ (N_0=0).
\end{align}
The first term $\rho^2_{\rm in}$ of Eq.~(\ref{rho.2frag}) gives a measure of the sum of 
the squared matter radii of the fragments (each $\rho_i^2/N_i$ is the mean-square matter radius of the $i$th fragment), 
while the second term $\rho^2_{\rm rel}$ is exactly the same as that of Eq.~(\ref{N-body}) with $A_i=N_i$, giving a measure of the spatial extension of the relative motion of the fragments. It is natural to arrange the coordinates into cluster-internal and cluster-relative to describe 
the motion of the $K$ fragments. The cluster-internal coordinates, denoted by 
$(\bm x_1, \bm x_2, \ldots, \bm x_{N-K})$, consist of a collection of Jacobi coordinates of each fragment, and the cluster-relative coordinates denoted by 
$(\bm x_{N-K+1}, \bm x_{N-K+2}, \ldots, \bm x_{N-1})$ are 
Jacobi coordinates as defined by Eq.~(\ref{XtoR}). Clearly $\rho^2$ 
is independent of the number of fragments into which the $N$-nucleon system develops.

\section{Equation of motion in adiabatic hyperspherical expansion}
\label{eq.Motion}

To solve a Schr\"odinger equation for the system in the hyperspherical method, 
a total wave function $\Psi$ is usually expanded in terms of a complete set of 
the hyperspherical harmonics or $K$-harmonics $Y_{\lambda}(\Omega)$: $\Psi=\rho^{-(d-1)/2}\sum_{\lambda}\chi_{\lambda}(\rho)Y_{\lambda}(\Omega)$~\cite{barnea01,timofeyuk08,cavagnero86,barnea97}. Here $Y_{\lambda}(\Omega)$ is an eigenfunction of 
$T_{\Omega}$ labeled by $\lambda$. 
The hyperradial functions $\chi_{\lambda}(\rho)$ are determined from a set of coupled-channels equations. This 
method is successfully used in nuclear few-body systems~\cite{zhukov93}. However, 
the number of hyperspherical harmonics needed to reach a converged solution becomes very large at large $\rho$ values~\cite{descouvemont10}. 
Moreover, the coupling matrix elements between different $Y_{\lambda}(\Omega)$ 
are of the same orders of magnitude as the diagonal matrix elements especially for the Coulomb interaction, which also makes the convergence slow. 

In low-energy phenomena, the hyperradial motion is expected to be slow compared to 
the hyperangular motion. Thus an adiabatic 
potential that takes account of all possible hyperangular motion at a fixed hyperradius 
gives insight into the dynamics of the system's evolution~\cite{fano81}. 
We adopt the adiabatic hyperspherical expansion method~\cite{lin95,nielsen01,mehta09,rakshit12,daily14} used extensively in atomic and molecular physics. 
%With that approach the many-body Schr\"odinger equation is recast 
%to such a form that describes the large-scale change of the system in a consistent manner. 
 
We define an adiabatic Hamiltonian $H_{\rm ad}$ by   
\begin{align}
H_{\rm ad} &= T_{\Omega}+V+\frac{\hbar^2 (d-3)(d-1)}{8m \rho^2}
=H-T_{\rho}+\frac{\hbar^2 (d-3)(d-1)}{8m \rho^2}.
\label{adiabatic}
\end{align} 
Here $V$ is the total potential energy and $H=T+V$ is the total Hamiltonian of the system.  
The nucleon-nucleon interaction of $V$ is assumed to be 
an effective interaction that contains no strong short-ranged repulsion. Assuming that $V$ contains no derivative of $\rho$, we solve an eigenvalue problem of the Hermitian operator $H_{\rm ad}$,
\begin{align}
H_{\rm ad}\Phi_{\nu}(\rho, \Omega)=U_{\nu}(\rho)\Phi_{\nu}(\rho, \Omega),
\label{coll.H}
\end{align}
to obtain channel wave functions $\Phi_{\nu}(\rho, \Omega)$ and real adiabatic potentials 
$U_{\nu}(\rho)$ that are labeled by $\nu$. The quantum numbers of 
$H$ such as spin-parity $J^{\pi}$ are preserved as those of $H_{\rm ad}$, and the antisymmetry requirement on $\Psi$ applies on $\Phi_{\nu}(\rho, \Omega)$ as well. Note that $\rho$ appears parametrically in Eq.~(\ref{coll.H}). 
At fixed $\rho$, all possible couplings among various hyperangular configurations are taken into account to obtain $U_{\nu}(\rho)$ and $\Phi_{\nu}(\rho, \Omega)$. The $\Phi_{\nu}(\rho, \Omega)$ form a set of orthonormal functions at each  $\rho$, 
\begin{align}
\langle \Phi_{\nu'}(\rho, \Omega) | \Phi_{\nu}(\rho, \Omega) \rangle_{\Omega}=\delta_{\nu, \nu'},
\label{orthonormal}
\end{align} 
where $\langle \ldots \rangle_{\Omega}$ indicates that the integration is 
carried out over $\Omega$ with $\rho$ being fixed. Actually, the $\Phi_{\nu}(\rho, \Omega)$ 
also contain spin and isospin coordinates that have to be integrated, but they are 
omitted for the sake of simplicity. Apparently $U_{\nu}(\rho)$ contain 
the minimum `centrifugal potential', ${\hbar^2(d-3)(d-1)}/{8m\rho^2}$, 
for $N \geq 3$ even when the eigenvalue of ${\cal K}(\Omega)^2$ vanishes. 

The Schr\"odinger equation, $H\Psi=E\Psi$, 
is solved by expanding $\Psi$ in terms of $\Phi_{\nu}(\rho, \Omega)$: 
\begin{align}
\Psi=\rho^{-(d-1)/2}\sum_{\nu}f_{\nu}(\rho)\Phi_{\nu}(\rho, \Omega).
\end{align}
The normalization of $\Psi$ is $\sum_{\nu}\int_0^{\infty}|f_{\nu}(\rho)|^2 d\rho=1$ for a bound state. The hyperradial functions $f_{\nu}(\rho)$ are determined by solving a set of coupled-channels equations, 
\begin{align}
&\left[ -\frac{\hbar^2}{2m}\frac{d^2}{d \rho^2}+U_{\nu}(\rho)-E \right]f_{\nu}(\rho)
%\notag \\
%&\ -\frac{\hbar^2}{2m}\sum_{\nu'}\left[ 2P_{\nu \nu'}(\rho)\frac{d}{d \rho}+Q_{\nu \nu'}(\rho)\right]f_{\nu'}(\rho)=0
-\frac{\hbar^2}{2m}\sum_{\nu'}\left[ 2P_{\nu \nu'}(\rho)\frac{d}{d \rho}+Q_{\nu \nu'}(\rho)\right]f_{\nu'}(\rho)=0,
\label{cceq}
\end{align}
with non-adiabatic coupling terms
\begin{align}
&P_{\nu \nu'}(\rho)=\langle \Phi_{\nu}(\rho, \Omega) | \frac{\partial}{\partial \rho} \Phi_{\nu'}(\rho, \Omega) \rangle_{\Omega}, \ \ \ \ \  Q_{\nu \nu'}(\rho)=\langle \Phi_{\nu}(\rho, \Omega) | \frac{\partial^2}{\partial \rho^2} \Phi_{\nu'}(\rho, \Omega) \rangle_{\Omega}.
\label{ad.coupling}
\end{align}
Equations~(\ref{coll.H}), (\ref{cceq}), and (\ref{ad.coupling}) give  
a microscopic description of the large-scale dynamics. 

A unique advantage of the adiabatic hyperspherical approach is that both lower and upper 
bounds to the exact lowest energy of $H$ are readily obtained~\cite{starace79,coelho91}. As shown in Appendix~\ref{app.b}, we have 
\begin{align}
P_{\nu \nu}(\rho)=0.
\label{pnunu}
\end{align}
Its differentiation 
with respect to  $\rho$ leads to 
\begin{align}
\bar{Q}_{\nu \nu}(\rho) +Q_{\nu \nu}(\rho)=0,
\end{align}
where 
\begin{align}
\bar{Q}_{\nu \nu}(\rho)= \langle \frac{\partial}{\partial \rho}\Phi_{\nu}(\rho, \Omega) | \frac{\partial}{\partial \rho} \Phi_{\nu}(\rho, \Omega) \rangle_{\Omega}
\end{align}
is non-negative, and consequently $Q_{\nu \nu}(\rho) \leq 0$. 
The potential defined by  
\begin{align}
W_{\nu}(\rho)=U_{\nu}(\rho)-\frac{\hbar^2}{2m}Q_{\nu \nu}(\rho)
\label{WtoU}
\end{align}
always satisfies $W_{\nu}(\rho) \geq U_{\nu}(\rho)$.  
The lowest eigenvalue $E$ obtained by truncating Eq.~(\ref{cceq}) to a single-channel  
equation with the lowest adiabatic potential $U_0(\rho)$ or $W_0(\rho)$ gives a lower or 
upper bound to the exact lowest energy of $H$. See Appendix~\ref{app.b} for details. 
Convergence of the solution of Eq.~(\ref{cceq}) is checked by increasing the number $\nu$ of channels.

A time-dependent Schr\"odinger equation is convenient for studying how final 
configurations in e.g. few-body decay and sub-barrier fusion evolve from their initial states. 
The wave function at time $t$ is assumed as 
\begin{align}
\Psi(t)=\rho^{-(d-1)/2}\sum_{\nu}f_{\nu}(\rho, t)\Phi_{\nu}(\rho, \Omega).
\end{align}
Once $f_{\nu}(\rho, 0)$ are given, 
$f_{\nu}(\rho, t)$ for $t>0$ are determined from the equation
\begin{align}
&i\hbar \frac{\partial}{\partial t}f_{\nu}(\rho, t)=\left[ -\frac{\hbar^2}{2m}\frac{\partial^2}{\partial \rho^2}+U_{\nu}(\rho) \right]f_{\nu}(\rho, t)
%\notag \\
%&\ -\frac{\hbar^2}{2m}\sum_{\nu'}\left[ 2P_{\nu \nu'}(\rho)\frac{\partial}{\partial \rho}+Q_{\nu \nu'}(\rho)\right]f_{\nu'}(\rho, t).
-\frac{\hbar^2}{2m}\sum_{\nu'}\left[ 2P_{\nu \nu'}(\rho)\frac{\partial}{\partial \rho}+Q_{\nu \nu'}(\rho)\right]f_{\nu'}(\rho, t).
\end{align}

\section{Hyperradius dependence of adiabatic potential}
\label{rho-dep.pot}
 
The $\rho$-dependence of $U_{\nu}(\rho)$ or $W_{\nu}(\rho)$ governs 
how the nucleus responds to its change of size. 
The kinetic energy and the Coulomb potential respectively 
give $1/\rho^2$ and $1/\rho$ contributions to $U_{\nu}(\rho)$ at large $\rho$ values. 
Short-range pairwise nuclear interactions give a $\rho^{-n}  (n\geq 3)$ 
contribution~\cite{thompson00}.  
Let us focus on the lowest adiabatic potential with the same spin-parity $J^{\pi}$ as that of the ground state. $U_{0}(\rho)$ has a minimum at
$\rho \approx \rho_{\rm min}$ corresponding to the matter size of the ground state. 
As $\rho$ decreases from $\rho_{\rm min}$, $U_{0}(\rho)$ rises because of a 
loss of nuclear potential energy as well as an 
increase in the kinetic energy. As $\rho$ increases from $\rho_{\rm min}$, 
various configurations contribute to determining $\Phi_{0}(\rho, \Omega)$. 
Here, deformations, shell effects, couplings with different modes and so on participate in determining 
the adiabatic potential. $U_{0}(\rho)$ reaches a peak at some $\rho$ value or may even have 
a couple of local peaks at different $\rho$ values.  
As $\rho$ increases further, $U_{0}(\rho)$ approaches the lowest decay threshold of the nucleus. 

The above global feature of the adiabatic potential well corresponds to 
the decomposition of $\rho^2$ in conformity with a formation of fragments or clusters. 
As shown in Eq.~(\ref{rho.2frag}), the different decomposition of the fragments 
can be treated on an equal footing in the hyperspherical approach, 
which makes it possible to assess 
what configurations play an important role in determining the adiabatic potential. 
If one instead calculates a sort of adiabatic potential or potential energy surface 
as a function of the 
relative distance between two fragments, there is no way to compare such 
potentials for different fragment decompositions because their relative distances 
have a different meaning.  

What fragment decompositions or configurations contribute to the lowest adiabatic 
potential clearly depends on $\rho$. The expectation value of $H$ 
is a major contribution to the adiabatic potential (see Eq.~(\ref{adiabatic})). We 
rewrite $H$ according to the fragment decomposition:  
\begin{align}
H=\sum_{i=1}^K (T_i+V_i) + T_{\rm rel}+ V_{\rm rel},
\label{decomp.H}
\end{align}
where $T_i+V_i$ is the intrinsic Hamiltonian of the $i$th fragment, $T_{\rm rel}$  
the kinetic energy of the relative motion among the fragments, and 
$V_{\rm rel}$ denotes the potential energies acting between the nucleons 
belonging to the different fragments. 
$V_{\rm rel}$ depends on both cluster-internal and cluster-relative coordinates, 
thus causing a coupling of the relative motion among the fragments with their 
intrinsic motion.    
When $\rho_{\rm rel}$ is so large compared to $\rho_{\rm in}$ that  
the nucleon-nucleon interactions of $V_{\rm rel}$ can 
be neglected and only the leading term of the Coulomb potentials of $V_{\rm rel}$ is retained, $V_{\rm rel}$ reduces to   
\begin{align}
V_{\rm rel} \to V_{\rm rel}^C&=\sum_{j>i=1}^K \frac{Z_iZ_je^2}{|\bm R_i-\bm R_j|}
=\frac{e^2}{\rho_{\rm rel}}C(\Omega_{\rm rel}),
\label{coul.contr}
\end{align}
where $Z_ie$ is the charge of the $i$th fragment and $\rho_{\rm rel}$ and $\Omega_{\rm rel}$ 
are the hyperradius and hyperangles constructed from the cluster-relative coordinates 
$(\bm x_{N-K+1}, \bm x_{N-K+2}, \ldots, \bm x_{N-1})$. 
With increasing $\rho$ the intrinsic motion of each fragment is stabilized 
toward its own ground state, while the configurations responsible for 
the relative motion are decoupled from the intrinsic motion. Both the coupling and 
decoupling of various degrees of freedom are naturally taken into account in the hyperspherical approach.

\begin{figure}
\begin{center}
\epsfig{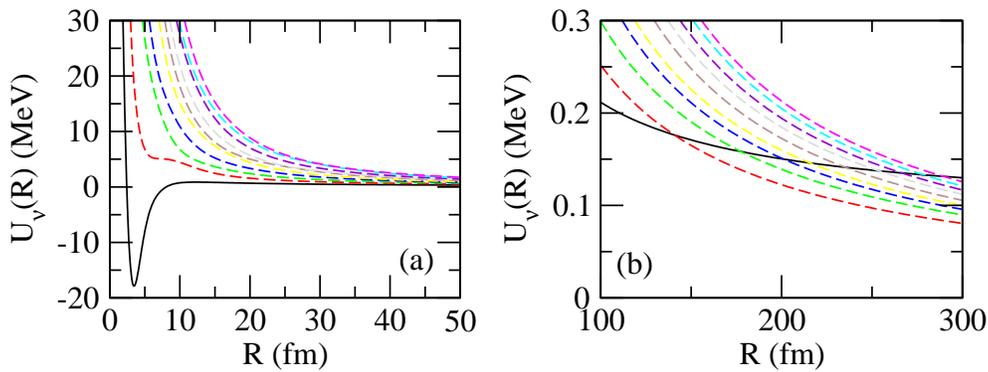}
\caption{The 10 lowest adiabatic potential curves of the three-$\alpha$ system with  $J^{\pi}=0^+$ that are taken from Ref.~\cite{suno15}. The hyperradius denoted $R$ here is defined by $R^2=\sqrt{3}\sum_{i=1}^3(\bm R_i-\bm R_{\rm cm})^2$, and the energy is measured from the three-$\alpha$ threshold. The solid line 
denotes the adiabatic potential dominated by the $^8$Be+$\alpha$ channel, while the dashed lines denote the 
potentials dominated by the three-$\alpha$ continuum channel. Panel (b) is an enlarged 
view of the potentials at large $R$ where a number of sharp avoided crossings successively appear.}
\label{fig.triplealpha}
\end{center}
\end{figure}

When there are several thresholds corresponding to
different fragment decompositions, avoided crossings of the adiabatic potential 
energy curves may occur. As an example, we show the case of $^{12}$C that 
is described with a cluster model of three $\alpha$-particles~\cite{suno15}. 
The eigenvalue problem~(\ref{coll.H}) for $H_{\rm ad}$ is solved 
accurately, and an analysis of the adiabatic potentials clarifies how 
the contributions of the hyperangular kinetic energy, the nuclear potential and 
the Coulomb potential change as a function of $\rho$.  Figure~\ref{fig.triplealpha}, taken from Fig. 2 of Ref.~\cite{suno15}, displays the 10 lowest adiabatic potential curves for $J^{\pi}=0^+$. The lowest potential $U_0(R)$ has a minimum at $R\approx 3.5$\,fm, which is deep enough to support a bound state, that is, the ground state of $^{12}$C. Furthermore, the 
lowest potential reaches a broad peak around 
12\ fm, corresponding to the second $0^+$ state of $^{12}$C, the Hoyle resonance state. 
The adiabatic potential indicated by the solid line is dominated by the 
two-body $^8$Be+$\alpha$ state and approaches the $^8$Be+$\alpha$ threshold at large $R$, while the other potentials indicated by the dashed lines 
are all dominated by the three-$\alpha$ continuum states. 
As seen in Fig.~\ref{fig.triplealpha} (b), an avoided crossing begins to occur at $R\approx 140$\,fm, which is because the three-$\alpha$ continuum state comes down closely to the two-body $^8$Be+$\alpha$ state. Since the avoided crossing actually occurs within a small range of $R$, it may be hard to see it in the figure. Refer to Fig. 3 of 
Ref.~\cite{suno15} to confirm the crossing clearly. Since the $^8$Be+$\alpha$ threshold is  
higher than the three-$\alpha$ threshold, a number of avoided crossings successively 
appear below the adiabatic potential indicated by the solid line. 
As is well known, the non-adiabatic coupling terms~(\ref{ad.coupling}) may be 
singular especially when the avoided crossing is sharp, namely it occurs within a small range of $\rho$. In that case, a diabatic procedure is proposed for accurately solving 
Eq.~(\ref{cceq})~\cite{lin95,tolstikhin96,suno11}. 
The slow variable discretization method combined with a complex absorbing potential makes it possible to solve Eq.~(\ref{cceq}) and to   
reproduce the energy and width of the Hoyle resonance in good agreement with experiment~\cite{suno15}.

\begin{figure}
\begin{center}
\epsfig{file=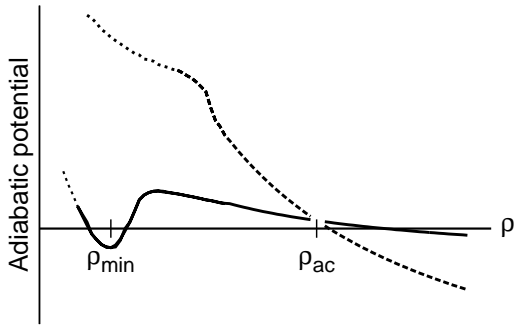,scale=1.1}
\caption{A schematic diagram of the adiabatic potential curves of 
${}_{\ 98}^{252}$Cf as a function of hyperradius $\rho$. Energy is measured from the 
ground state of ${}_{\ 98}^{252}$Cf. The potential dominated by the $\alpha+{}_{\ 96}^{248}$Cm 
channel (solid line) goes to 
$-6.2$ MeV at large $\rho$ values, while that by the 
${}_{\ 54}^{140}{\rm Xe}+{}_{\ 44}^{108}{\rm Ru}+4n$ channel (dashed line) to $-200.4$ MeV. An avoided crossing occurs at $\rho \approx \rho_{\rm ac}$.  
The lowest adiabatic potential $U_0(\rho)$ changes its dominant character from $\alpha+{}_{\ 96}^{248}$Cm to the ${}_{\ 54}^{140}{\rm Xe}+{}_{\ 44}^{108}{\rm Ru}+4n$ channel around $\rho_{\rm ac}$. }
\label{fig.Cfdecay}
\end{center}
\end{figure}

Let us speculate concerning the adiabatic potential curves of ${}_{\ 98}^{252}$Cf that are crucially important for determining its decay mode. 
The ground state of $^{252}$Cf decays mostly by an $\alpha$-particle emission. 
The rest is a spontaneous fission (SF), emitting 3.7 neutrons on average. 
To make things simple, we approximate the SF as occurring through 
a single channel of ${}_{\ 54}^{140}{\rm Xe}+{}_{\ 44}^{108}{\rm Ru}+4n$. 
The two decay modes contain different numbers of fragments, 
two in $\alpha+{}_{\ 96}^{248}$Cm and six in the SF, but  
the hyperspherical approach can treat both in a unified way. 
The threshold of $\alpha+^{248}$Cm is 6.2\,MeV below the ground 
state of $^{252}$Cf, whereas that of the SF is 200.4 MeV lower than 
the ground state. See the schematic diagram of Fig.~\ref{fig.Cfdecay}. The lowest adiabatic potential $U_{0}(\rho)$ 
approaches the SF threshold at large $\rho$. Above that threshold 
many $U_{\nu}(\rho)$ curves, not drawn in Fig.~\ref{fig.Cfdecay}, show up 
corresponding to the continuum states of the SF mode. A unique  
$U_{\nu}(\rho)$ with the two-body $\alpha+^{248}$Cm character appears 
high above the SF threshold. When moving inward from this asymptotic region, the Coulomb potential (\ref{coul.contr}) produces a distinct difference between the two decay modes.  The charge factor $Z_1Z_2$ of the SF mode is more than ten times larger than that 
of the $\alpha$ channel. Thus those $U_{\nu}(\rho)$ curves that are dominantly contributed  by the SF configurations rise up rapidly,  
while the $U_{\nu}(\rho)$ curve of the $\alpha$ channel increases much more slowly. 
At the avoided crossing point $\rho_{\rm ac}$, the lowest curve $U_{0}(\rho)$ comes very close to that of the $\alpha$ curve, and for  
$\rho < \rho_{\rm ac}$ the $\alpha$ channel makes a dominant contribution to 
$U_{0}(\rho)$. With further decrease of $\rho$ many different configurations begin to 
mix due to an increasing role of the nuclear interaction $V$. The $U_{0}(\rho)$ reaches 
a barrier top around some point and reaches its minimum at $\rho_{\rm min}$ corresponding to the matter radius of the ground state of $^{252}$Cf. Though much more complicated than the $^{12}$C case, the gross feature of the adiabatic potential curves of ${}_{\ 98}^{252}$Cf 
should have some similarity to those of $^{12}$C, and the decay branch of ${}_{\ 98}^{252}$Cf will be determined by solving Eq.~(\ref{cceq}).

\section{Multicluster approximation and integration over hyperangles}
\label{Matrix.element}

Solving Eq.~(\ref{coll.H}) is of vital importance in the adiabatic hyperspherical approach.  
Its accurate solution is obviously very hard except for few-body 
system. The difficulty is enhanced by the fact that the matrix element has to be calculated by integrating over 
$\Omega$ only. Some efforts have been made for extending to larger 
systems~\cite{timofeyuk08,rakshit12,daily14}. We take up this problem assuming the use of 
many-body wave functions that contain all the coordinates. 

Before discussing the eigenvalue problem~(\ref{coll.H}), we note 
that a usual approach defines an adiabatic potential barrier or energy surface at a 
given `collective' coordinate by searching for a minimum of $V$ 
for various parameters that characterize the nuclear density or shape~\cite{brack72}. 
This makes sense in that $V$ is a major part of $H_{\rm ad}$, and because, since $V$ is a 
function of $\rho$ and $\Omega$, its minimum gives information on the most important 
$\Omega$ values contributing to the lowest adiabatic potential. As mentioned before, 
the adiabatic hyperspherical approach can go beyond that by taking account of 
various couplings with different degrees of freedom.

Let us assume that the channel wave function $\Phi_{\nu}(\rho, \Omega)$ at a given $\rho$ is expanded in terms of suitable basis functions $\phi_i(\bm x)$:
\begin{align}
\Phi_{\nu}(\rho, \Omega)=\sum_i C_{\nu i}(\rho)\phi_i(\bm x).
\label{channelwf}
\end{align} 
Equation~(\ref{coll.H}) is then reduced to the following generalized eigenvalue equation 
for determining the coefficients $C_{\nu i}(\rho)$ and the adiabatic potential $U_{\nu}(\rho)$: 
\begin{align}
\sum_j [{\cal H}_{ij}(\rho)-U_{\nu}(\rho){\cal B}_{ij}(\rho)]C_{\nu j}(\rho)=0,
\end{align}
where ${\cal H}_{ij}(\rho)$ and ${\cal B}_{ij}(\rho)$ are adiabatic Hamiltonian and overlap matrices defined by 
\begin{align}
{\cal H}_{ij}(\rho)=\langle \phi_i (\bm x)|H_{\rm ad} |\phi_j(\bm x)\rangle_{\Omega},\ \ \ \ \ 
{\cal B}_{ij}(\rho)=\langle \phi_i (\bm x) |\phi_j(\bm x)\rangle_{\Omega}.
\label{matrixHandB}
\end{align} 
We include only those basis functions that give 
a $c$-number $\rho^2$ for the expectation value of the squared hyperradius operator $\tilde{\bm x} \bm x$:    
\begin{align}
\frac{\langle  \phi_i(\bm x)|\tilde{\bm x} \bm x|\phi_i(\bm x) \rangle}{\langle  \phi_i(\bm x)|\phi_i(\bm x) \rangle}\approx \rho^2. 
\label{constraint}
\end{align}

We face two problems. One is what basis functions we use for $\phi_i(\bm x)$. 
The other is how to calculate the matrix element in Eq.~(\ref{matrixHandB}). The first 
one is crucially important for assessing the quality of $\Phi_{\nu}(\rho, \Omega)$ and 
$U_{\nu}(\rho)$. Though it is difficult to give a general answer, 
our ansatz is to employ a microscopic multicluster approximation~\cite{varga94,arai01}.   
This is because, as mentioned in Sects.~\ref{intro} and \ref{rho-dep.pot}, the structure 
change we are interested in includes a variety of configurations ranging from one-centered shell-model wave functions to those with a few fragments or subsystems. A general form of the multicluster wave function containing $K$ fragments reads  
\begin{align}
\phi^{(K)}(\bm x)={\cal A}\{\Psi_1(\bm z_1)\Psi_2(\bm z_2)\cdots\Psi_K(\bm z_K)\chi(\bm x_{N-K+1},\ldots,\bm x_{N-1})\},
\label{multicluster}
\end{align} 
where ${\cal A}$ is an antisymmetrizer, $\Psi_i$ an antisymmetrized intrinsic state of the $i$th fragment containing 
$N_i$ nucleons and $\chi$ is the relative motion function for the fragments. 
The cluster-internal coordinates 
$(\bm x_1,\bm x_2,\ldots,\bm x_{N-K})$ are abbreviated as 
$(\bm z_1, \bm z_2,\ldots,\bm z_K)$, where e.g. $\bm z_1$ stands for the first $N_1-1$ Jacobi 
coordinates $(\bm x_1,\bm x_2,\ldots,\bm x_{N_1-1})$. The spin-isospin coordinates are again 
suppressed. In general $\Psi_i$ may represent 
not only the ground state of the fragment 
but also its excited state. The quantum numbers for characterizing $\Psi_i$ are omitted. 
The coupling of the     
angular momenta of the $\Psi_i$s and $\chi$ to a total angular momentum $JM$ is  
implicitly understood in Eq.~(\ref{multicluster}). 
We presume $\phi_i(\bm x)$ to belong to the space spanned by 
\begin{align}
\{\phi^{(1)}(\bm x)\}+\{\phi^{(2)}(\bm x)\}+\ldots
\end{align}
Note that any states in $\{\phi^{(K)}(\bm x)\}$ are in general nonorthogonal to each other 
even when they belong to different $K$ subspaces. The questions of what intrinsic states of the fragments are important and what $K$ subspaces have to be included 
depend on a given system and energy range of interest. To proceed further, we 
assume that $\Psi_i$ is approximated by harmonic-oscillator 
shell-model wave functions, while $\chi$ is described well 
with a superposition of Gauss functions~\cite{varga95,book,hiyama03} as developed in few-body problems. 

We have to calculate a matrix element for some operator $O(\bm x)$, 
\begin{align}
{\cal O}(\rho_0)=\langle \phi_i (\bm x)|O(\bm x) |\phi_j(\bm x)\rangle_{\Omega, \,\rho=\rho_0},
\label{o.rho}
\end{align}
by integrating over $\Omega$ at fixed $\rho$, say $\rho_0$. 
The  calculation of the matrix element of $T_{\rho}$ 
in $H_{\rm ad}$ can be aided with use of the identity 
\begin{align}
\frac{\partial }{\partial \rho} \phi_j(\bm x)=\frac{1}{\rho}\Big(\sum_{i=1}^{N-1}
\bm x_i\cdot \frac{\partial}{\partial \bm x_i}\Big) \phi_j(\bm x).
\end{align} 
See Appendix~\ref{cghyp} for an example. In calculating the matrix element of $H$, 
the cluster-intrinsic term $\sum_{i=1}^K(T_i+V_i)$ (see Eq.~(\ref{decomp.H})) may be replaced by 
\begin{align}
H\phi^{(K)}(\bm x) &\to \Big(\sum_{i=1}^KE_i\Big)\phi^{(K)}(\bm x) \notag \\
&+ 
{\cal A}\{(T_{\rm rel}+V_{\rm rel})\Psi_1(\bm z_1)\Psi_2(\bm z_2)\cdots\Psi_K(\bm z_K)\chi(\bm x_{N-K+1},\ldots,\bm x_{N-1})\},
\end{align}
using the observed energy $E_i$ of $\Psi_i$. 
This approximation looks reasonable and practically useful 
because any nuclear interaction can not 
satisfactorily reproduce the saturation property of nuclear binding energies despite 
the fact that reproducing the 
threshold energy for the fragment decomposition is important in the present approach. 

The second problem has so far been examined using integral transform 
techniques~\cite{baz70,daily14}. We use a $\delta$ function technique 
as in Ref.~\cite{daily14}. 
Using the expression for Dirac $\delta$ function
\begin{align}
\delta(\rho-\rho_0)=\frac{1}{\pi \rho_0}\int_{-\infty}^{\infty}\,e^{i\omega(1-\rho^2/\rho_0^2)}d\omega,
\end{align}
we can express ${\cal O}(\rho_0)$ as a Fourier transform of $F_{\rho_0}(\omega)$:  
\begin{align}
{\cal O}(\rho_0)&=\frac{1}{\pi} \int_{-\infty}^{\infty} e^{i \omega} F_{\rho_0}(\omega) d\omega, 
\label{ftr.vs.o}
\\
F_{\rho_0}(\omega)&=\int e^{-i\omega \tilde{\bm \xi}\bm \xi} \big(\phi_i (\rho_0\bm \xi)\big)^* O(\rho_0\bm \xi) \phi_j(\rho_0\bm \xi) d\bm \xi.
\label{frho.omega}
\end{align}
Note that $\bm x_i$ is changed to $\rho_0\bm \xi_i$ 
with a dimensionless variable $\bm \xi_i$. In Eq.~(\ref{frho.omega})
$d\bm \xi$ stands for $d\bm \xi_1d\bm \xi_2 \ldots d\bm \xi_{N-1}$, where the integration 
range of each $\bm \xi_i$ covers the whole three-dimensional space.   
Since $e^{-i\omega \tilde{\bm \xi}\bm \xi}=\prod_{k=1}^{N-1}e^{-i\omega \bm \xi_k^2}$ results in a 
simple modification of the basis function, $F_{\rho_0}(\omega)$ can be calculated 
with a technique developed in microscopic cluster models~\cite{horiuchi77,slyv03}. 

In some cases the Fourier integral~(\ref{ftr.vs.o}) can easily be obtained by 
Cauchy's integral formula 
that reduces to a residue calculation. Whether or not we have a practical means for evaluating Eq.~(\ref{o.rho}) for a general case 
depends on how fast and accurately the Fourier integral is computed. 
For this aim we test the Whittaker cardinal series or the Whittaker-Shannon interpolation formula~\cite{mcnamee71}: 
\begin{align}
F_{\rho_0}(\omega)=\sum_{n=-\infty}^{\infty}F_{\rho_0}(\omega_n){\rm sinc} {\textstyle{\frac{\pi}{h}}}(\omega-\omega_n),
\label{sampl.rep}
\end{align}
where sinc\,$x$ is the sinc function, ${\sin x}/{x}$, and $\omega_n=nh\, (n=0,\pm1,\pm2,\ldots)$ is the grid of the sampling points. The series~(\ref{sampl.rep}) is known to converge if $F_{\rho_0}(\omega)$ is a band-limited function.
Because  sinc\,$n\pi =\delta_{n,0}$, the series is exact at 
all the sampling points. It is in fact an expansion in terms of 
orthogonal functions $\{{\rm sinc} {\textstyle{\frac{\pi}{h}}}(\omega-\omega_n)\}$ that have the properties:  
\begin{align}
&\int_{-\infty}^{\infty} {\rm sinc} {\textstyle{\frac{\pi}{h}}}(\omega-\omega_n) d\omega =h, \notag \\
&\int_{-\infty}^{\infty} {\rm sinc} {\textstyle{\frac{\pi}{h}}}(\omega-\omega_m) 
\,{\rm sinc} {\textstyle{\frac{\pi}{h}}}(\omega-\omega_n) d\omega =h\delta_{m,n},\notag \\
&\int_{-\infty}^{\infty} e^{i\omega} {\rm sinc} {\textstyle{\frac{\pi}{h}}}(\omega-\omega_n) d\omega
=he^{i\omega_n}\ \ \ ({\textstyle{\frac{h}{\pi}}}<1).
\end{align} 
The third equation called the Dirichlet integral leads to an approximation for 
$O(\rho_0)$:   
\begin{align}
O(\rho_0) \approx \frac{h}{\pi} \sum_{n=-M}^{M} 
F_{\rho_0}(\omega_n)\, e^{i\omega_n}, 
\label{approx}
\end{align}
which is nothing but a trapezoidal rule for the integration. This result is  
due to the fact that the Fourier transform of the sinc 
function is the rectangular function and vice versa. 
To determine $M$, we need to know how fast $F_{\rho_0}(\omega)$ decreases as a function of $\omega$. The mesh size $h\, (h < \pi)$ is determined by examining how accurate the expansion is 
at, e.g. $\omega=(n+\frac{1}{2})h$, the midpoint of $\omega_n$ and $\omega_{n+1}$. 

Other interpolations, e.g. a cubic spline interpolation may also be worthwhile testing because it 
leads to a simple expression for Eq.~(\ref{ftr.vs.o}) and in addition the mesh size can be taken as piecewise variable. Once $d F_{\rho_0}(\omega)/d\omega$ values at  
both boundaries of the interpolation are calculated, we can completely fix the interpolating function of the cubic spline. 

Since $\omega$-dependence of $F_{\rho_0}(\omega)$ is of practical importance, we examine 
it for the diagonal matrix elements ($\phi_i(\bm x)=\phi_j(\bm x)$) of ${\cal O}(\bm x)=1$ and $T_{\Omega}$ in a very schematic model. As the model, we 
employ CG ignoring the antisymmetry requirement of the 
wave function and focus only on its spatial part.  See 
%Refs.~\cite{stecher09,rakshit12,daily14} for its application to the hyperspherical approach and 
Appendix~\ref{cghyp} for some basic matrix elements with the CG. 
For a spherical CG, $\exp(-{\textstyle{\frac{1}{2}}}\tilde{\bm x}A\bm x)$, 
the positive-definite symmetric matrix $A$ is set to 
${\rm Tr} A^{-1} = \frac{2}{3} \rho_0^2$ 
because of Eqs.~(\ref{constraint}) and (\ref{exp.val.rhorho}). 
We may choose $A$ to be diagonal, $A=(a_{i}\delta_{ij})$, as 
far as the diagonal matrix element of $O(\rho_0)$ is concerned. 

Our first choice for $A$ is 
a uniform nuclear expansion, $a_{i}=a$, leading to a hyperscalar Gaussian,  $\exp(-{\textstyle{\frac{1}{2}}}\tilde{\bm x}A\bm x)=
\exp(- {\textstyle{\frac{1}{2}}}a \rho^2)$. 
This function is totally symmetric and $\Omega$-independent. By taking $a$ as 
$(N-1)/{a}=\frac{2}{3}\rho^2_0$, the overlap matrix element is (see Eq.~(\ref{F.ovl}))
\begin{align}
F_{\rho_0}(\omega)=\left(\frac{(2\pi)^{N-1}}{(2a\rho_0^2+2i\omega)^{N-1}}\right)^{3/2}=\left(\frac{\pi}{{d}/{2}+i\omega}\right)^{d/2}.
\end{align}
Clearly $|F_{\rho_0}(\omega)|$ becomes very small if $\omega$ is significantly 
larger than $d/2$. The Fourier 
transform~(\ref{ftr.vs.o}) can be rigorously computed in this case. If $d/2$ is an integer, $F_{\rho_0}(\omega)$ has a pole 
of order $d/2$ at $\omega=id/2$, so that the integral is reduced to a residue calculation, 
yielding 
\begin{align}
O(\rho_0)=\frac{2\sqrt{\pi}^d}{\Gamma(d/2)}e^{-d/2}.
\label{spherical.gauss}
\end{align}
Even when $d/2$ is a half integer, we can derive the above result as follows. 
By the change of the integration variable, $d/2+i\omega=-t$, $O(\rho_0)$ is reduced to 
\begin{align}
O(\rho_0)=\frac{i\sqrt{\pi}^d}{\pi} e^{-d/2} 
\int_{-d/2+i\infty}^{-d/2-i\infty} e^{-t}(-t)^{-d/2} dt.
\end{align}
By changing the integration path to the Hankel contour and using Hankel's integral  representation and Euler's reflection formula for the gamma function, we find the above integral to be $2\pi/i\Gamma(d/2)$. 
The result~(\ref{spherical.gauss}) is in fact trivial thanks to 
Eq.~(\ref{surface.integral}) if we note that the hyperscalar 
Gaussian at $\rho=\rho_0$ is $\exp(- {\textstyle{\frac{1}{2}}}a \rho_0^2)
=e^{-d/4}$ and hence $O(\rho_0)$ must be $e^{-d/2}\int d\Omega$. 
We note that $O(\rho_0)$ for ${\cal O}(\bm x)=T_{\Omega}$ 
vanishes because the hyperscalar Gaussian is $\Omega$-independent and, when acted on by $T_{\Omega}$, vanishes. This is also confirmed by using Eq.~(\ref{T.omega.g}).

\begin{figure}
\begin{center}
\epsfig{file=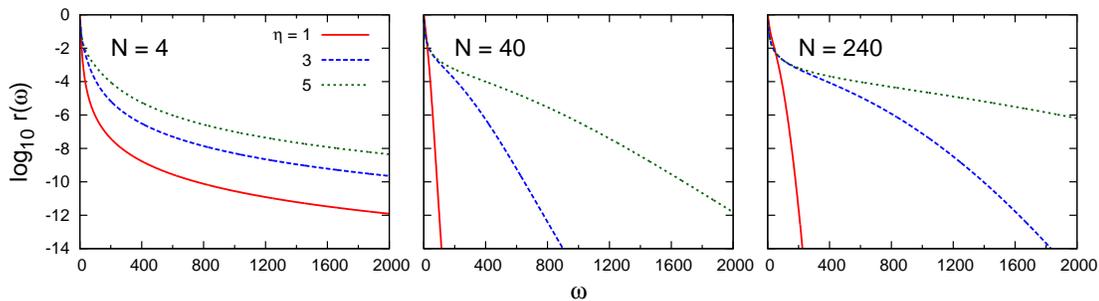,scale=0.9}
\caption{The ratio  $r(\omega)=|F_{\rho_0}(\omega)/F_{\rho_0}(0)|$ for the overlap corresponding to 
$N$ nucleons' symmetric fission as a function of $\omega$. $\rho_0$ 
is set to $\eta \sqrt{N}R_{\rm rms}(N)$: 
Solid, dashed, and dotted lines correspond to $\eta=1, 3$, and 5, respectively. See text for details.}
\label{fig.ovlp}
\end{center}
\end{figure}

\begin{figure}
\begin{center}
\epsfig{file=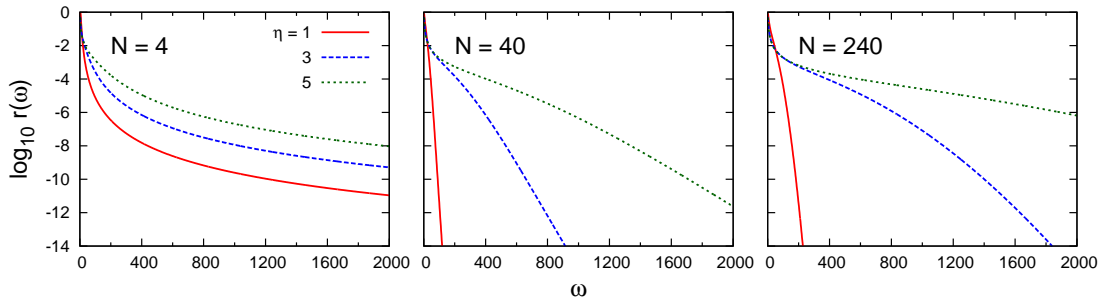,scale=0.9}
\caption{ The same as Fig.~\ref{fig.ovlp} but for the hyperangular kinetic energy.}
\label{fig.kine}
\end{center}
\end{figure}

The next example is a `symmetric fission', that is,  
the nucleus fissions into two identical  fragments  with mass number $N/2$ and only the relative motion between them expands with increasing $\rho_0$. Let $R_{\rm rms}(N)$ denote the root-mean-square radius of a nucleus with mass number $N$, and set it equal to $\sqrt{3/5}r_0N^{1/3}$ ($r_0=1.1$ fm). The matrix $A$ for the symmetric fission is chosen as $a_1=a_2=\ldots=a_{N-2}=a$, and $a$ and $a_{N-1}$ are determined by the condition
\begin{align}
\frac{N-2}{a}+\frac{1}{a_{N-1}}= \frac{2}{3} \rho_0^2,\ \ \ \ \ \ \frac{N/2-1}{a}= \frac{2}{3} \rho_f^2,
\end{align}
where $\rho_f$ is fixed to $\sqrt{N/2}R_{\rm rms}(N/2)$. The mass number $N$ is changed 
to 4, 40, and 240, and for each $N$ $\rho_0$ is taken 
as $\rho_0=\eta \sqrt{N}R_{\rm rms}(N)$ ($\eta=1,3, 5$). 
Figure~\ref{fig.ovlp} displays $\log_{10}r(\omega)$ for the overlap, where $r(\omega)=|F_{\rho_0}(\omega)/F_{\rho_0}(0)|$.  
In the case of $N=4$, the fall-off of $r(\omega)$ 
is slow with increasing $\omega$ and $\eta$. For example, $\log_{10} r(\omega)$ at $\omega=5000$ is $-13.7,\, -11.4,\, -10.1$ for $\eta=1, 3, 5$, respectively. 
For $N=40$, $r(\omega)$ rapidly drops 
to $10^{-15}$ as a function of $\omega$, but its decrease becomes slower for $\eta=5$. 
This behavior is also valid for $N=240$, and the decrease in $\omega$ becomes even 
slower with increasing $\eta$.  As shown in Fig.~\ref{fig.kine}, the ratio 
$\log_{10} r(\omega)$ for $T_{\Omega}$ is very similar to that of the overlap.

\section{Evolution of intrinsic shapes} 
\label{sec.shape}

It is interesting to know how an intrinsic shape of the nucleus changes with increasing $\rho$. When a decay or an SF  
is considered as a tunneling through a barrier,  
the shape will give insight into where the fragments are formed and how they evolve  
during the passing through the barrier. The barrier is conventionally 
calculated by assuming some density distribution constrained with shape or deformation parameters such as quadrupole and octupole~\cite{brack72,krappe12}. Such deformations are not observable, however. Our view is to reverse this approach.  
Since the nucleus should in principle preserve the total angular momentum, it is not trivial 
to imagine the intrinsic shape in the space-fixed frame. For example, any state with $L=0$ is spherical in that frame, but it can happen that such state is intrinsically deformed and rotates. 
As shown in Ref.~\cite{wiringa00}, the intrinsic two-$\alpha$ structure of the rotational state of $^8$Be emerges from  
the wave function obtained by a quantum Monte Carlo calculation. 

Following the procedure of Ref.~\cite{wiringa00}, we can get the intrinsic density or deformation indicated by e.g. the lowest channel wave function $\Phi_{0}(\rho, \Omega)$.  Since $\Phi_{0}(\rho, \Omega)$ is normalized as in 
Eq.~(\ref{orthonormal}), its square, 
$P_{\rho}(\Omega)=|\Phi_{0}(\rho, \Omega)|^2$, gives the probability density as a function of $\Omega$ at a given $\rho$.  First, we generate many sampling points 
$\Omega_{1}, \Omega_{ 2}, \ldots, \Omega_{M}$ according 
to the distribution of $P_{\rho}(\Omega)$ using the Metropolis-Hastings algorithm. 
Secondly, we define a body-fixed intrinsic frame for each  
$\Omega_{j}=(\bm \xi^{j}_1, \bm \xi^{j}_2,\ldots,\bm \xi^{j}_{N-1})$ as follows. 
By using Eq.~(\ref{r.to.x}) together with $\bm R_{\rm cm}=0$, 
Jacobi coordinates $\bm x^j_i=\rho\bm \xi^j_i\, (i=1,2, \ldots, N-1)$ specify 
the positions of $N$ nucleons $(\bm r^j_1, 
\bm r^j_2, \ldots, \bm r^j_N)$ in the space-fixed frame.  From these position vectors, 
we calculate the moment of inertia tensor
\begin{align}
{\cal I}^j_{\alpha \beta}
=\sum_{i=1}^{N}  {{r_i^j}_{\alpha}}{{r_i^j}_{\beta}}, 
\end{align} 
where ${{r_i^j}_{\alpha}}$ ($\alpha=x,y,z$) is the Cartesian 
component of 
$\bm r_i^{j}$.  Diagonalizing the $3\times 3$ symmetric matrix ${\cal I}^j$ determines the principal moments of inertia, which define the axes of the intrinsic frame. For example, 
the axis is called $x', y', z'$ in increasing order of the principal 
moment of inertia. The direction of the axis also has to be chosen consistently.  
By reading $(\bm r^j_1, \bm r^j_2, \ldots, \bm r^j_N)$ as 
$({{\bm r}'}^{j}_1, {{\bm r}'}^{j}_2,\ldots, {{\bm r}'}^{j}_N)$ 
in reference to the intrinsic frame, we obtain the desired position coordinates of $N$ nucleons in the intrinsic frame. 
Finally, accumulating these position coordinates over $j=1,2, \ldots, M$ leads to  
the intrinsic single-particle density 
at $\rho$. Once the intrinsic density is obtained, it is easy to  
extract multipole deformations.

\section{Eigenvalue problem of Hamiltonian with radius constraint}
\label{sec.constraint}

It looks as though the adiabatic hyperspherical approach has some relationship to an eigenvalue problem of the Hamiltonian with a constraint~\cite{ring80}. Let us attempt to find a 
solution of the Schr\"odinger equation by first constraining the expectation value of the 
squared hyperradius to a fixed $\rho^2$ value. Suppose that the solution is expanded in terms of some basis functions:
\begin{align}
\Psi_{\rho \kappa}(\bm x)=\sum_{i} G_{\kappa i}(\rho) \phi_i(\bm x).
\end{align}
The constraint~(\ref{constraint}) is not necessarily imposed on $\phi_i(\bm x)$ itself, but  
we demand the solution we are looking for to satisfy the condition
\begin{align}
\frac{\langle \Psi_{\rho \kappa}(\bm x)|\tilde{\bm x}\bm x|\Psi_{\rho \kappa}(\bm x)\rangle}{\langle \Psi_{\rho \kappa}(\bm x)|\Psi_{\rho \kappa}(\bm x)\rangle}=\rho^2.
\label{Q.constraint}
\end{align}
Here $\kappa$ is a label related to the Lagrange multiplier. The unknown coefficients 
$G_{\kappa i}(\rho)$ and the energy eigenvalue $E_{\kappa}(\rho)$ are determined from the following equation 
\begin{align}
\sum_j \left[H_{ij}-\kappa(Q_{ij}-\rho^2 B_{ij})-E_{\kappa}(\rho)B_{ij}\right] 
G_{\kappa j}(\rho)=0,
\label{const.Sch.eq}
\end{align}
where $H$, $B$, and $Q$ are matrices defined by 
\begin{align}
{H}_{ij}=\langle \phi_i (\bm x)|H|\phi_j(\bm x)\rangle,\ \ \ \ \ 
{B}_{ij}=\langle \phi_i (\bm x) |\phi_j(\bm x)\rangle,\ \ \ \ \ 
{Q}_{ij}=\langle \phi_i (\bm x) |\tilde{\bm x}\bm x|\phi_j(\bm x)\rangle.
\label{matrixH.B.Q}
\end{align}
Unlike Eq.~(\ref{matrixHandB}), the above matrices are obtained by integrating over the 
whole coordinates. To determine the coefficients $G_{\kappa j}(\rho)$ from 
Eq.~(\ref{const.Sch.eq}), the value of $\kappa$ has to be given. Actually 
$\kappa$ should be such that both Eqs.~(\ref{Q.constraint}) and (\ref{const.Sch.eq}) are simultaneously met. Apparently $\Psi_{\rho' \kappa'}(\bm x)$ and $\Psi_{\rho \kappa}(\bm x)$ are not orthogonal to each other even for $\rho'=\rho$.   

The next step is to use the generator coordinate method in which a solution $\Psi$ for the  
Schr\"odinger equation is assumed as
\begin{align}
\Psi=\sum_{\kappa}\int C_{\kappa}(\rho)\Psi_{\rho \kappa}(\bm x) d\rho.
\end{align} 
The coefficients $C_{\kappa}(\rho)$ are determined from the Hill-Wheeler equation
\begin{align}
\sum_{\kappa}\int \langle \Psi_{\rho' \kappa'}(\bm x)|H-E|\Psi_{\rho \kappa}(\bm x)\rangle
C_{\kappa}(\rho)d\rho=0,
\end{align}
which should be satisfied for any  $\rho'$ and $\kappa'$ values. 
An approximate solution to the Hill-Wheeler 
equation gives an upper bound to the ground-state energy. Note that the adiabatic hyperspherical approach gives both lower and upper bounds as discussed in 
Sect.~\ref{eq.Motion}.

We refer to two interesting calculations with a constraint in comparison to the adiabatic hyperspherical approach. One is  
a Hartree-Fock-Bogoliubov calculation performed by constraining 
the mean-square radius, $\sum_{i=1}^N\bm r_i^2/N$, to study how self-conjugate nuclei fragment into $\alpha$ 
clusters~\cite{girod13}. As Eq.~(\ref{rho2.xx}) indicates, this constraint is equivalent 
to that of $\rho^2$ provided the contribution of $\bm R_{\rm cm}^2$ to the squared radius 
remains a constant. The treatment of the 
cm motion in Ref.~\cite{girod13} does not satisfy this condition as usual in a mean-filed 
model. It would be a challenge for the mean-field approximation to cope with such diverse 
structure at large distances that is composed of different numbers of fragments.  
What should be further pursued at this moment is to establish the essential 
relationship between the adiabatic hyperspherical approach and `beyond mean-field' calculations or configuration interaction calculations that constrain the mean-square matter radius. 

Another is a simultaneous study of both $\alpha+^6$He reactions and the 
structure change of $^{10}$Be in a microscopic $\alpha+\alpha+n+n$ model~\cite{ito06}, 
in which a distance parameter between the two $\alpha$-clusters is 
constrained. Since the motion of the two neutrons is restricted to either molecular  
or atomic orbits around the $\alpha$-clusters, the main configurations included 
are $\alpha+^6$He and $^5$He+$^5$He two-body types. The adiabatic 
energy surfaces are calculated within that approximation. An avoided crossing is treated by the generator coordinate method. 
As noted in Sect.~\ref{rho-dep.pot}, the relative distance of the fragments is not a collective coordinate. If one constrains $\rho^2$ as the generator coordinate, it would be 
possible in the same four-body model to take account of possible couplings with 
the $^9$Be+$n$ channel that 
is the lowest threshold of $^{10}$Be as well as the three- and four-body channels, 
$^8$Be+$n+n$ and $\alpha+\alpha+n+n$, that are open in the energy region treated 
in Ref.~\cite{ito06}.

\section{Conclusion}
\label{summary}

Stressing that the hyperradius is a collective coordinate, we 
have formulated a fully microscopic adiabatic hyperspherical approach 
to large-scale nuclear dynamics. The equation of motion for 
hyperradial functions is universal, independent of the number of nucleons, and 
enables one to consistently treat the dynamics from confined nuclear motion to relative motion among fragments in their asymptotic region. It is possible to describe 
in a unified way cases where the nucleus fragments into several channels.
No spurious center-of-mass motion appears and couplings 
with different degrees of freedom can naturally be taken into account. These properties are 
due to the fact that both the squared hyperradius and the kinetic energy are  
flexibly decomposed into cluster-internal and cluster-relative 
quantities responding to the fragment formation.  

The adiabatic potential as a function of the hyperradius plays a key role in the present approach. It is unambiguously defined solely by the Hamiltonian of the system, and 
there is no need to assume specific geometrical shapes or deformations to compute it. 
Conversely the shape or intrinsic density, if necessary, comes out after the adiabatic potential  
is obtained or the equation of motion for the hyperradial functions is solved. The calculation of the adiabatic potential 
involves the integration over all the coordinates but the hyperradius. 
Expecting that a microscopic multicluster model is a promising 
candidate for applying the present approach to larger systems, we 
have discussed the use of Fourier transforms for evaluating the matrix elements needed to 
obtain the adiabatic potential. The matrix elements can be 
obtained in exactly the same way as the usual matrix elements needed in 
nuclear many-body calculations. A merit of the Fourier transform technique 
is its simplicity, and test calculations indicate that accurate evaluations of the 
matrix elements are feasible. 

Although the calculation of the adiabatic potential still requires much 
computer time for large systems, 
a real challenge is whether we can provide large enough basis functions to cover important configurations for fixed $\rho$. 
Further developments are certainly indispensable for a microscopic, realistic description of  large-scale nuclear dynamics.

\section*{Acknowledgment}

The author is greatly indebted to H. Suno for many instructive discussions. He also thanks W. Horiuchi, K.~M. Daily, and C.~H. Greene  for valuable communications. 
This work is supported in part by JSPS KAKENHI Grant No. 24540261. 


%\bibliographystyle{ptephy}
%\bibliography{sample}
%
% once the .bbl file has been generated then place the text in your article.

\appendix

\section{Lower and upper bounds}
\label{app.b}
In this appendix we rigorously prove Eq.~(\ref{pnunu}) and show that both lower and upper bounds to the ground-state energy are respectively obtained by solving single-channel equations. 

The ground-state wave function may be expressed in the hyperspherical coordinates as
\begin{align}
\Psi=\rho^{-(d-1)/2}f(\rho)\Phi(\rho,\Omega)
\end{align}
with the normalization condition 
\begin{align}
\int_0^{\infty}|f(\rho)|^2 d\rho=1,\ \ \ \ \ \ \langle \Phi(\rho, \Omega)|\Phi(\rho, \Omega)\rangle_{\Omega}=1.
\end{align}
The hyperradial function $f(\rho)$ has to vanish at $\rho=0$. 
The ground-state energy reads  
\begin{align}
E_{\rm exact}
&=-\frac{\hbar^2}{2m}\int_0^{\infty}f^*(\rho)\Big[ \frac{d^2 f(\rho)}{d\rho^2}+2\frac{d f(\rho)}{d\rho} P(\rho) 
%\notag \\
%&\ \ \ +f(\rho) Q(\rho)\Big]d\rho  + \int_0^{\infty}|f(\rho)|^2 U(\rho)d\rho,
+f(\rho) Q(\rho)\Big]d\rho  + \int_0^{\infty}|f(\rho)|^2 U(\rho)d\rho,
\label{Exact.E}
\end{align}
where
\begin{align}
&P(\rho)=\langle \Phi(\rho,\Omega)|\frac{\partial}{\partial \rho} \Phi(\rho,\omega)\rangle_{\Omega},\ \ \ \ \ \ 
Q(\rho)=\langle\Phi(\rho,\Omega)|\frac{\partial^2}{\partial \rho^2} \Phi(\rho,\Omega)\rangle_{\Omega}, \notag \\
&U(\rho)=\langle \Phi(\rho,\Omega)|H_{\rm ad}|\Phi(\rho,\Omega)\rangle_{\Omega}.
\label{def.PQU}
\end{align}
From the normalization condition of $\Phi(\rho,\Omega)$, we obtain
\begin{align}
\frac{d}{d \rho} 
\langle\Phi(\rho,\Omega)|\Phi(\rho,\Omega)\rangle_{\Omega}
=P(\rho)^* +P(\rho)=0.
\end{align}
Thus $P(\rho)$ must be pure imaginary or zero. 
If $P(\rho)$ is not zero but pure imaginary, 
$f^*(\rho)df(\rho)/d\rho$ in Eq.~(\ref{Exact.E}) must also be 
pure imaginary because $E_{\rm exact}$ is real. With  $f(\rho)=g(\rho)+ih(\rho)$, where 
$g(\rho)$ and $h(\rho)$ are real functions, 
$f^*(\rho)df(\rho)/d\rho$ reads 
\begin{align}
f^*(\rho)\frac{df(\rho)}{d\rho}
&=\frac{1}{2}\frac{d}{d\rho}\{g(\rho)^2+h(\rho)^2\} 
%\notag \\
%&+i\{g(\rho)h'(\rho)-h(\rho)g'(\rho)\},
+i\{g(\rho)h'(\rho)-h(\rho)g'(\rho)\},
\end{align}
which leads to  $d\{g(\rho)^2+h(\rho)^2\}/d\rho=0$. 
Thus $g(\rho)^2+h(\rho)^2$ is a constant, and it 
must be zero because of $f(0)=0$. Namely, $f(\rho)$ vanishes identically, which can not  
be accepted. Using $P(\rho)=0$ in Eq.~(\ref{Exact.E}) leads to  
\begin{align}
E_{\rm exact}
&=\int_0^{\infty}f^*(\rho)\left(-\frac{\hbar^2}{2m}\frac{d^2}{d\rho^2}+W(\rho)\right)f(\rho)d\rho
\label{exact.energy}
\end{align}
with
\begin{align}
W(\rho)&=U(\rho)-\frac{\hbar^2}{2m}Q(\rho).
\end{align} 

Suppose that for $\Phi(\rho, \Omega)$ we take the $\Phi_0(\rho, \Omega)$ that gives the lowest  adiabatic potential. The corresponding quantities 
$Q(\rho)$ and $U(\rho)$ in Eq.~(\ref{def.PQU}) are denoted by $Q_0(\rho)$ and $U_0(\rho)$, 
respectively. It follows from the Ritz variational principle that 
\begin{align}
E_{\rm exact} \leq \int_0^{\infty}f^*(\rho)\left(-\frac{\hbar^2}{2m}\frac{d^2}{d\rho^2}+W_0(\rho)\right)f(\rho) d\rho.
\end{align}
If $f(\rho)$ is chosen to be the solution of the equation (the adiabatic approximation), 
\begin{align}
\left(-\frac{\hbar^2}{2m}\frac{d^2}{d\rho^2}+W_0(\rho)\right)f(\rho) =E_{\rm U}f(\rho),
\label{eq.ub}
\end{align}
with the lowest eigenvalue $E_{\rm U}$, $E_{\rm U}$ turns out to be an 
upper bound of $E_{\rm exact}$:   
$E_{\rm exact} \leq E_{\rm U}$. 
Differentiating $P(\rho)=0$ with respect to $\rho$ leads to  
\begin{align}
\langle \frac{\partial}{\partial \rho}\Phi(\rho,\Omega)|\frac{\partial}{\partial \rho}\Phi(\rho,\Omega)\rangle_{\Omega} 
+ Q(\rho)
=0.
\end{align}
Equation~(\ref{exact.energy}) for $E_{\rm exact}$ is recast to    
\begin{align}
E_{\rm exact}
&=\int_0^{\infty}f^*(\rho)\Big[-\frac{\hbar^2}{2m}\frac{d^2}{d\rho^2}+ U(\rho) 
%\notag \\ 
%& +\frac{\hbar^2}{2m}\langle\frac{\partial }{\partial \rho}\Phi(\rho,\Omega)|\frac{\partial}{\partial \rho}\Phi(\rho,\Omega)\rangle_{\Omega}\Big]f(\rho) d\rho.
+\frac{\hbar^2}{2m}\langle\frac{\partial }{\partial \rho}\Phi(\rho,\Omega)|\frac{\partial}{\partial \rho}\Phi(\rho,\Omega)\rangle_{\Omega}\Big]f(\rho) d\rho.
\end{align}
Since the last term in the square brackets is non-negative, 
we obtain 
\begin{align}
E_{\rm exact} & \geq 
\int_0^{\infty}f^*(\rho)\left(-\frac{\hbar^2}{2m}\frac{d^2}{d\rho^2}+U(\rho)\right)f(\rho) d\rho. 
\end{align}
By using the inequality $U(\rho)\geq U_0(\rho)$ and choosing $f(\rho)$ to be 
the solution of the equation (the Born-Oppenheimer approximation),  
\begin{align}
\left(-\frac{\hbar^2}{2m}\frac{d^2}{d\rho^2}+U_0(\rho)\right)f(\rho) =E_{\rm L}f(\rho),
\label{eq.lb}
\end{align} 
with the lowest eigenvalue $E_{\rm L}$, we obtain a lower bound of $E_{\rm exact}$ as 
$E_{\rm exact} \geq E_{\rm L}$.

If we calculate the expectation value of $H$ for the wave function 
$\Psi=\rho^{-(d-1)/2}f(\rho)\Phi_{\nu}(\rho,\Omega)$ with the $\nu$th channel wave 
function, we confirm Eq.~(\ref{pnunu}) using the same argument as above.

\section{Matrix elements with correlated Gaussians}
\label{cghyp}

In this appendix we calculate $F_{\rho_0}(\omega)$, Eq.~(\ref{frho.omega}), 
using as $\phi_i(\bm x)$ the generating function 
$g(\bm s; A, \bm x)$~\cite{varga95,book,suzuki08,aoyama12} of the CG: 
\begin{align}
g(\bm s; A, \bm x)=\exp(-{\textstyle{\frac{1}{2}}}\tilde{\bm x}A\bm x+\tilde{\bm s}\bm x),
\label{gfn.cg}
\end{align}
where $A$ is an $(N-1)\times (N-1)$ symmetric, positive-definite matrix and 
$\bm s=(\bm s_i)$ is an $(N-1)$-dimensional column vector to describe motion with 
non-zero orbital angular momentum. They are both parameters that characterize the CG. 
The constraint~(\ref{constraint}) reads  
\begin{align}
\frac{3}{2}{\rm Tr} A^{-1}+ \tilde{\bm s}A^{-2} \bm s \approx \rho_0^2.
\label{exp.val.rhorho}
\end{align}
Note that for the special case that $A$ is diagonal, $A=(a_i\delta_{i,j})$, 
$g(\bm s; A, \bm x)$ reduces to a product of Gaussian wave packets: 
$g(\bm s; A, \bm x)=\prod_{i=1}^{N-1} \exp [ -\frac{1}{2}a_i
(\bm x_i-\bm s_i)^2+\frac{1}{2}a_i\bm s_i^2 ]$. 
We present formulas for $F_{\rho_0}(\omega)$ calculated between  
$g(\bm s; A, \bm x)$ and $g(\bm s'; A', \bm x)$.  See Ref.~\cite{book} for details. 
%The parameters $A'$ and $\bm s'$ also satisfies the condition~(\ref{exp.val.rhorho}). 
The case with $\bm s=\bm s'=0$ is given in Ref.~\cite{stecher09}. \\   

%\noindent
{\it Overlap}
\par\noindent
The function $F_{\rho_0}(\omega)$ for ${\cal O}(\bm x)=1$ is given by
\begin{align}
F_{\rho_0}(\omega)
&=\left(\frac{(2\pi)^{N-1}}{{\rm det}B}\right)^{3/2}e^{-\frac{1}{2}\tilde{\bm v}B^{-1}\bm v}
\label{F.ovl}
\end{align}
with
\begin{align}
B=\rho_0^2(A+A')+2i\omega I,\ \ \ \ \ \bm v=\rho_0 (\bm s+\bm s').
\end{align}
Here $I$ is the $(N-1)\times (N-1)$ identity matrix. Since $A+A'$ can be diagonalized by 
an orthogonal matrix, the matrix $B$ can be diagonalized as well. \\ 

%\noindent
{\it Kinetic energy}
\par\noindent
To calculate the matrix element for ${\cal O}(\bm x)=T_{\Omega}=T-T_{\rho}$, we use 
the following relation~\cite{stecher09}:
\begin{align}
\frac{\partial}{\partial \rho}g(\bm s; A, \bm x)
&=\frac{1}{\rho}(-\tilde{\bm x}A\bm x+\tilde{\bm s}\bm x)g(\bm s; A, \bm x),\notag \\
\frac{\partial^2}{\partial \rho^2}g(\bm s; A, \bm x)&=\frac{1}{\rho^2}\Big[
(\tilde{\bm x}A\bm x)^2-\tilde{\bm x}A\bm x+(\tilde{\bm s}\bm x)^2 %\notag \\
%&\quad -2(\tilde{\bm x}A\bm x)\tilde{\bm s}\bm x\Big]g(\bm s; A, \bm x),\notag \\
-2(\tilde{\bm x}A\bm x)\tilde{\bm s}\bm x\Big]g(\bm s; A, \bm x),\notag \\
Tg(\bm s; A, \bm x)
&=-\frac{\hbar^2}{2m}\Big[-3{\rm Tr}A+\tilde{\bm s}\bm s -2\tilde{\bm s}A\bm x 
%\notag \\
%&\qquad +\tilde{\bm x}A^2\bm x\Big]g(\bm s; A, \bm x).
+\tilde{\bm x}A^2\bm x\Big]g(\bm s; A, \bm x).
\end{align}
Combining these results, we obtain 
\begin{align}
%&T_{\Omega}g(\bm s; A, \bm x)\notag \\
%&=\frac{\hbar^2}{2m\rho^2}\Big[3\rho^2{\rm Tr}A-\rho^2\tilde{\bm s}\bm s+2\rho^2\tilde{\bm s}A\bm x + (d-1)\tilde{\bm s}\bm x \notag \\
%&\qquad +(\tilde{\bm s}\bm x)^2-\rho^2\tilde{\bm x}A^2\bm x - d\tilde{\bm x}A\bm x \notag \\
%&\qquad -2(\tilde{\bm x}A\bm x)\tilde{\bm s}\bm x + (\tilde{\bm x}A\bm x)^2 \Big]g(\bm s; A, \bm x).
T_{\Omega}g(\bm s; A, \bm x)
&=\frac{\hbar^2}{2m\rho^2}\Big[3\rho^2{\rm Tr}A-\rho^2\tilde{\bm s}\bm s+2\rho^2\tilde{\bm s}A\bm x + (d-1)\tilde{\bm s}\bm x \notag \\
&+(\tilde{\bm s}\bm x)^2-\rho^2\tilde{\bm x}A^2\bm x - d\tilde{\bm x}A\bm x 
-2(\tilde{\bm x}A\bm x)\tilde{\bm s}\bm x + (\tilde{\bm x}A\bm x)^2 \Big]g(\bm s; A, \bm x).
\label{T.omega.g}
\end{align}
Here $d$ is defined in Eq.~(\ref{dimension}). 
The change of variables, $\bm x \to \rho_0 \bm \xi$, yields $F_{\rho_0}(\omega)$ as 
\begin{align}
%F_{\rho_0}(\omega)&=\frac{\hbar^2}{2m\rho^2} \int e^{
% -\frac{1}{2}\tilde{\bm \xi}B\bm \xi+\tilde{\bm v}\bm \xi } \notag \\
%&\times \Big[3\rho^2{\rm Tr}A-\rho^2\tilde{\bm s}\bm s+2\rho^3\tilde{\bm s}A\bm \xi + (d-1)\rho \tilde{\bm s}\bm \xi \notag \\
%&  +\rho^2(\tilde{\bm s}\bm \xi)^2-\rho^4\tilde{\bm \xi}A^2\bm \xi - d\rho^2\tilde{\bm \xi}A\bm \xi \notag \\
%&  -2\rho^3(\tilde{\bm \xi}A\bm \xi)\tilde{\bm s}\bm \xi + \rho^4(\tilde{\bm \xi}A\bm \xi)^2 \Big]d\bm \xi.
F_{\rho_0}(\omega)&=\frac{\hbar^2}{2m\rho_0^2} \int e^{
 -\frac{1}{2}\tilde{\bm \xi}B\bm \xi+\tilde{\bm v}\bm \xi } 
 \Big[3\rho_0^2{\rm Tr}A-\rho_0^2\tilde{\bm s}\bm s+2\rho_0^3\tilde{\bm s}A\bm \xi + (d-1)\rho_0 \tilde{\bm s}\bm \xi \notag \\
&  +\rho_0^2(\tilde{\bm s}\bm \xi)^2-\rho_0^4\tilde{\bm \xi}A^2\bm \xi - d\rho_0^2\tilde{\bm \xi}A\bm \xi 
  -2\rho_0^3(\tilde{\bm \xi}A\bm \xi)\tilde{\bm s}\bm \xi + \rho_0^4(\tilde{\bm \xi}A\bm \xi)^2 \Big]d\bm \xi.
\end{align}
This integral can be performed analytically.  
In the case of $\bm s=\bm s'=0$, we obtain 
\begin{align}
%F_{\rho_0}(\omega)&=\frac{\hbar^2}{2m\rho^2} \left(\frac{(2\pi)^{N-1}}{{\rm det}B}\right)^{3/2} 
% \Big[ 3{\rm Tr}\rho^2 A  \notag \\
%&\  -3 {\rm Tr}B^{-1}(\rho^2 A)^2-3d{\rm Tr}B^{-1}\rho^2 A \notag \\
%&\  +9({\rm Tr}B^{-1}\rho^2 A)^2+6{\rm Tr}(B^{-1}\rho^2 A)^2\Big].
F_{\rho_0}(\omega)&=\frac{\hbar^2}{2m\rho_0^2} \left(\frac{(2\pi)^{N-1}}{{\rm det}B}\right)^{3/2} 
 \Big[ 3{\rm Tr}\rho_0^2 A  
 -3 {\rm Tr}B^{-1}(\rho_0^2 A)^2-3d{\rm Tr}B^{-1}\rho_0^2 A \notag \\
&\  +15({\rm Tr}B^{-1}\rho_0^2 A)^2-12 M_2(B, \rho_0^2 A)\Big].
\label{F.T_omega}
\end{align}
Here use is made of the formula  
\begin{align}
\int e^{ -\frac{1}{2}\tilde{\bm x}B\bm x}(\tilde{\bm x} P\bm x)^2 d\bm x
=\left(\frac{(2\pi)^{N-1}}{{\rm det}B}\right)^{3/2} \Big[ 15({\rm Tr}B^{-1}P)^2-12 M_2(B,P)\Big]
\end{align}
for an $(N-1) \times (N-1)$ symmetric matrix $P$, where $M_2(B, P)$ is defined by 
\begin{align}
M_2(B, P)=\frac{1}{{\rm det}B}\sum_{j > i=1}^{N-1}
\left|
\begin{array}{cccc}
B_{11} & B_{12} & \ldots & B_{1\, N-1} \\
\vdots & \vdots & \vdots & \vdots      \\
P_{i1} & P_{i2} & \ldots & P_{i\, N-1} \\
\vdots & \vdots & \vdots & \vdots      \\
P_{j1} & P_{j2} & \ldots & P_{j\, N-1} \\
\vdots & \vdots & \vdots & \vdots      \\
B_{N-1\,1} & B_{N-1\,2} & \ldots & B_{N-1\, N-1} \\
\end{array}
\right|.
\end{align}
\\

%\noindent
{\it Potential energy}
\par\noindent 
The matrix element for ${\cal O}(\bm x)=V$ is conveniently calculated by expressing the distance vector of two nucleons as a combination of Jacobi coordinates 
\begin{align}
\bm r_i-\bm r_j=\tilde{\zeta}\bm x,
\end{align}
where $\zeta$ is an $(N-1)$-dimensional column vector determined by $i$ and $j$. 
For a Gauss potential, ${\cal O}(\bm x)=e^{-\tau^2 (\bm r_i-\bm r_j)^2}=e^{-\tau^2 \tilde{\bm x}\zeta \tilde{\zeta}\bm x}$, $F_{\rho_0}(\omega)$ reduces to that of the overlap. 
For $\bm s=\bm s'=0$, we obtain 
\begin{align}
F_{\rho_0}(\omega)
&=\left(\frac{(2\pi)^{N-1}}{{\rm det}(B+2\tau^2 \rho_0^2 \zeta \tilde{\zeta})}\right)^{3/2}
=\left(\frac{(2\pi)^{N-1}}{ (1+2\tau^2 \rho_0^2\tilde{\zeta} B^{-1}\zeta)\,{\rm det}B}\right)^{3/2}. 
\label{gauss.me}
\end{align}
In the last step Sherman-Morrison formula, ${\rm det}(B+c\zeta \tilde{\zeta})=(1+c\tilde{\zeta} B^{-1}\zeta)\,{\rm det}B$, is used, where $c$ is a constant. 

By comparing this result with Eq.~(\ref{F.ovl}) and by noting that $B^{-1}$ is constrained 
by the condition~(\ref{exp.val.rhorho}), 
we expect that the contribution of the potential energy $V$ to the adiabatic potential behaves as $(\rho/\tau)^{-3}$ for large $\rho$ values. The $\rho^{-3}$ dependence was found for 
a three-body system~\cite{thompson00}, but our result suggests that it is valid for 
many-body systems as well.

As an important application of Eq.~(\ref{gauss.me}), 
we calculate the matrix element for the Coulomb potential, 
${\cal O}(\bm x)=1/|\bm r_i-\bm r_j|$. Using  
\begin{align}
\frac{1}{|\bm r_i-\bm r_j|}=\frac{2}{\sqrt{\pi}}\int_0^{\infty}e^{-\tau^2 
(\bm r_i-\bm r_j)^2}d\tau,
\end{align}
and the integral 
\begin{align}
\int_0^{\infty}(1+a\tau^2)^{-3/2}d\tau=\frac{1}{\sqrt{a}},
\end{align}
we obtain the matrix element for the Coulomb potential as 
\begin{align}
F_{\rho_0}(\omega)&=\sqrt{\frac{2}{\pi}}\frac{1}{\rho_0}  
\left(\frac{(2\pi)^{N-1}}{{\rm det}B}\right)^{3/2} (\tilde{\zeta}B^{-1}\zeta)^{-1/2}.
\end{align}
As expected, the inverse $\rho$-dependence appears naturally.

\end{document}